\documentclass[12pt,twoside]{article}
\usepackage{fleqn,espcrc1}

\usepackage{graphicx}
\usepackage{epsfig}

\usepackage[figuresright]{rotating}

\newcommand{\s}{$\bar{s}s$}
\newcommand{\p}{$\bar{p}p$}
\newcommand{\f}{$\phi$}
\newcommand{\an}{annihilation}
\newcommand{\ap}{antiproton}

\newcommand{\om}{$\omega$}
\newcommand{\ten}{$f'_2 (1525) $}
\newcommand{\la}{$\Lambda$}
\newcommand{\al}{$\bar{\Lambda}$}

\newcommand{\AmS}{{\protect\the\textfont2
  A\kern-.1667em\lower.5ex\hbox{M}\kern-.125emS}}

\title{Polarized  strangeness in the nucleon }

\author{M.G.Sapozhnikov \address{Laboratory of Particle Physics,
Joint Institute for Nuclear Research,\\
P.O. Box 141980, Dubna, Russia}}

\begin{document}

\maketitle

\begin{abstract}
A large violation of the Okubo-Zweig-Iizuka rule was discovered in the annihilation of stopped
antiprotons. The explanation of these experimental data is discussed in the framework of
the model assumed that the nucleon strange sea quarks are
polarized.
\end{abstract}

\section{INTRODUCTION}

According to the naive quark models the proton wave function contains just
two $u$-quarks and one $d$-quark and
the role of the strange quarks in the nucleon seems to be marginal.
However there are experimental indications that the
$\bar{s}s$ pairs in the nucleon are responsible for the number of non-trivial effects.

        It was found that the magnitude of the strange quarks contribution
varies for  different nucleon matrix elements. Thus, the fraction of the nucleon
momentum carried by the strange quarks is
not large \cite{CCFR.95}, \cite{MRS.98}:
\begin{equation}
P_s= 4.6\%~ at~Q^2=20~GeV^2
\end{equation}

The contribution of the strange quarks to the proton electric
form factor is also quite small. The HAPPEX Collaboration measurements
allow to extract the combinations of strange electric and magnetic form factors at
$Q^2=0.48~ (GeV/c)^2$ \cite{Ani.99}
\begin{equation}
G^s_E + 0.39 G^s_M= 0.023\pm0.034(stat)\pm0.022(syst)\pm0.026
\end{equation}
(the last error is related with
uncertainties in the neutron electric form factor).

However
the contribution of the strange quarks in the nucleon mass may be
substantial, according to the analysis of the $\pi N$ data for
evaluation of the nucleon $\sigma$-term \cite{San.97} it is:
\begin{equation}
m_s<N|\bar{s}s|N> \sim 130~ MeV
\end{equation}

 The strange quarks contribution to the nucleon magnetic moment,
measured by the SAMPLE Collaboration \cite{Spa.00}, is not small

\begin{equation}
G^s_M(0.1~GeV^2) = (+0.61\pm0.17\pm0.21\pm0.19) \mu_N \label{gsm}
\end{equation}
(where $\mu_N$ is the nuclear magneton and
the last error is due to the uncertainty in the axial form factor $G^Z_A$ of $Z^0$ exchange)
with a positive sign contrary to the predictions of many
theoretical models.

Moreover, during the past decade
the EMC and successor experiments with polarized lepton beams and
nucleon targets \cite{Ada.97}
gave indication that the \s\ pairs in the nucleon
are polarized.

\begin{equation}
\Delta s \equiv \int\limits_{0}^{1} dx[ s_{\uparrow}(x) -
s_{\downarrow}(x) + \bar s_{\uparrow}(x)-
\bar s_{\downarrow}(x)] = - 0.10\pm0.02.
\end{equation}
The minus sign means that the strange quarks and antiquarks are polarized
negatively with respect to the direction of the nucleon spin.

        Experiments on elastic neutrino scattering \cite{Gar.93} have also
provided an indication that the intrinsic nucleon strangeness is negatively
polarized though within  large uncertaintes.
It was obtained that $\Delta s = -0.15 \pm 0.07$.

        Recent common analysis \cite{Chi.00} of the baryon magnetic moments
with the data of
the SAMPLE Collaboration (\ref{gsm}) leads to the conclusion that the
best fit of these data gives
$\Delta s = -0.19 \pm 0.08$.

        The lattice QCD calculations  also indicate
of negative polarization of strange quarks in proton: $\Delta s = -0.12 \pm 0.01$
\cite{Don.95} and $\Delta s = -0.109 \pm 0.030$ \cite{Fuk.95}.

        In \cite{Ell.95} it was proposed to use the
assumption about polarization of nucleon strangeness to
explain the large OZI violation seen in different reactions of
\p~\an~at rest. The model
was extended to the reaction $\bar p p \rightarrow
\Lambda \bar \Lambda$ in~\cite{Alb.95}, where arguments were given on the
basis of
chiral symmetry that the $\bar s s$ pair in the nucleon wave
function might be in the $^3$$P_0$ state. Also
the model of~\cite{Ell.95} was applied in \cite{Ek.95}
to make predictions for
$\Lambda$ longitudinal polarization in the target fragmentation
region for deep-inelastic lepton scattering.

        The main aim of my talk is to summarize the present status of
the polarized strangeness model in view of new experimental facts
(for  recent developments of the model, see \cite{Ell.99}).

\section{STRONG VIOLATION OF THE OZI RULE}

        The OZI rule \cite{OZI} forbids creation of \s~mesons in
the interaction of non-strange particles. The production of, for instance, \f~
meson is allowed only via presence of the light quark component in
the \f~ wave function. The amount of this component is determined
by the deviation of mixing angle of the vector nonet from the ideal mixing
angle, for the ideal mixing angle \f~ should be a pure \s~state. The OZI rule predicts
that in all hadron reactions the ratio between the cross sections of \f~ and
\om~ production $R(\phi/\omega)$
should be:
\begin{equation}
R(\phi/\omega) = 4.2\cdot10^{-3} f
\label{rfi}
\end{equation}
where $f$ is a ratio of phase spaces of the reactions.

This prediction  was tested
many times in experiments using different hadron beams.
The analysis~\cite{Nom.00} of the experiments collected in the
Durham reactions database  has shown that in $\pi N$
interactions
the weighted average ratio of cross sections of \f~ and \om~production
at different energies is
\begin{equation}
\bar{R}=\frac{\sigma(\pi N\to \phi X)}{\sigma(\pi N\to \omega X)}
= (3.30\pm0.34)\cdot10^{-3}
\label{eq:RV1}
\end{equation}
without attempting to make a phase-space correction.

The weighted average ratio of cross sections of \f~ and \om~production
at different energies
in nucleon-nucleon interactions is somewhat higher, but
still qualitatively similar to the OZI value (\ref{rfi}):
\begin{equation}
\bar{R}=\frac{\sigma(NN \to \phi X)}{\sigma(NN\to \omega X)}
= (12.78\pm0.34)\cdot10^{-3}
\label{eq:RV2}
\end{equation}

The corresponding value for
\ap~\an~ in flight is:
\begin{equation}
\bar{R}=\frac{\sigma(\bar{p}p \to \phi X)}{\sigma(\bar{p}p\to \omega X)}
 =  (14.55\pm1.92)\cdot10^{-3}
\label{flight}
\end{equation}

These experiments indicate that
the naive OZI rule for the vector meson production is
generally valid within 10$\%$
accuracy. This is not so bad for a heuristic model, bearing in mind that
the OZI prediction is based only
on the value of the mixing angle derived from meson masses, and
applied at different energies from 100 MeV till 100 GeV.

From the point of view of theory, it is realized that the OZI rule reflects
important feature of the hadron interactions -
suppression of the flavor mixing transitions.
It reflects the absence of the
processes with pure gluonic intermediate
states. The value of the flavor mixing is channel-dependent. It
is large for the
pseudoscalar and scalar channels. For other channels the OZI
rule is a nice approximation.
As it was discussed in \cite{Ven.90}, the OZI limit
of QCD is a more accurate approximation than the large $N_c$ limit,
or the quenched approximation, or the topological expansion
($N_c \to \infty$ at fixed
$N_f/N_c$).

\begin{figure}[htb]
\begin{center}
\includegraphics[width=10cm]{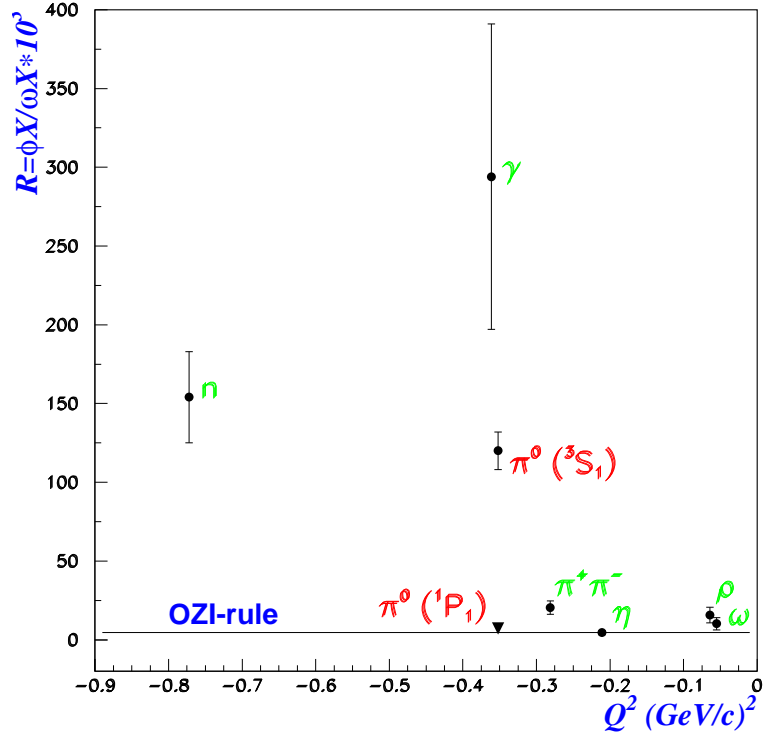}
\caption{The ratio $R=\phi X/\omega X \cdot 10^3$ of yields for different reactions of $\bar pp \to \phi (\omega) X$
annihilation at rest as
a function of the momentum transfer to \f~.The solid line shows the prediction of the OZI rule (\ref{rfi}).}
\label{ozi}
\end{center}
\end{figure}

In spite of this solid theoretical background and numerous experimental
confirmations there was a surprise when experiments at
LEAR (CERN) with stopped \ap s showed large violations of the OZI
rule (for a review, see  \cite{Nom.00,Ams.98,Sap.98}).
The compilation of the data is shown in Fig. \ref{ozi} where the ratio
$R=(\phi X/\omega X) \cdot 10^3$ of yields for different reactions of $\bar pp \to \phi (\omega) X$
annihilation at rest is shown as a
function of the momentum transfer to \f~. The solid line corresponds to the prediction of the OZI rule (\ref{rfi}).

Fig. \ref{ozi} demonstrates the following distinctive
features revealed by the LEAR experiments:

1) There is an unusually strong deviation from the OZI--rule
predictions.
Thus in the
$\bar{p} p \to \phi\gamma$
channel,  the Crystal Barrel collaboration has found~
\cite{Ams.98}, \cite{Ams.95} after phase space corrections:
\begin{equation}
R_{\gamma}={B(\bar p p \rightarrow \phi \gamma )
\over B( \bar p p \rightarrow \omega \gamma)}
= (294\pm 97)\cdot 10^{-3},
\label{Rgamma}
\end{equation}
which is about by 70 times larger than the
OZI prediction (\ref{rfi}).

Another very
large apparent violation of the OZI rule was found by the OBELIX and
Crystal Barrel collaborations in the
$ \bar{p} + p \to \phi(\omega) + \pi$ channel.

For the ratio of the phase space corrected branching ratios
 the Crystal Barrel measurement \cite{Ams.98} in liquid hydrogen gives:

\begin{equation}
R_\pi={B(\bar p p \rightarrow \phi \pi )
\over B( \bar p p \rightarrow \omega \pi)}= (106\pm 12)\cdot 10^{-3}
\label{Rpi}
\end{equation}

It coincides with the ratio of the \an~ yields measured by the
OBELIX Collaboration
for annihilation in a liquid-hydrogen target~\cite{Don.98}:

\begin{equation}
R_\pi = (114\pm 10)\cdot 10^{-3}
\label{Rpitwo}
\end{equation}

The ratios (\ref{Rpi}) and (\ref{Rpitwo}) are
about a factor of 30 higher than the OZI rule prediction.

2) The violation of the OZI-rule is not-universal for all \an~ channels of \f~ production but
mystically occurs only in some of them.
For instance, no enhancement of
\f~ production is observed for the $\phi \omega$
($R(\phi \omega/\omega \omega) = (19\pm 7)\cdot 10^{-3}$)
or $\phi\rho$
($R(\phi \rho/\omega \rho) = (6.3\pm 1.6)\cdot 10^{-3}$  \cite{Sap.98})
channels.

3) There is a strong dependence of the OZI--rule violation on
the quantum numbers of the initial \p~state.
It was clearly demonstrated by the OBELIX collaborations results:
\begin{eqnarray}
 R_{\pi}(\phi/\omega, ~^3S_1) & = & (120\pm 12)\cdot 10^{-3}~, \label{rfi1}\\
 R_{\pi}(\phi/\omega, ~^1P_1) & < & 7.2\cdot 10^{-3}
 ~~~~~~~~~~~~~~
 \mbox{, with~95\%~CL}~ \label{rfi2}
\end{eqnarray}

4) There is a serious indication
that the degree of the OZI rule violation depends on
the momentum transfer.

   To explain the huge violation of the OZI rule in the annihilation
of stopped antiprotons and its
strong dependence on the spin of the initial state,
the model
based on a nucleon wave function containing negatively polarized
$s \bar s$ pairs
was proposed
\cite{Ell.95}.

The model claims that the
observed OZI violation is only apparent because in these processes
the \s~meson is created via {\it connected} diagrams with participation
of intrinsic nucleon strange quarks.
The strong dependence on the
initial quantum numbers is due to polarization of the strange sea.
Let us discuss these assumptions in more details.

\section{Polarized strangeness model}

        Let us consider the production of \s~ strangeonia in
$NN$ or $\bar{N}N$ interactions assuming
that the nucleon wave function
contains an admixture of $\bar{s}s$ pairs which are polarized negatively
with respect to the direction of the nucleon spin.

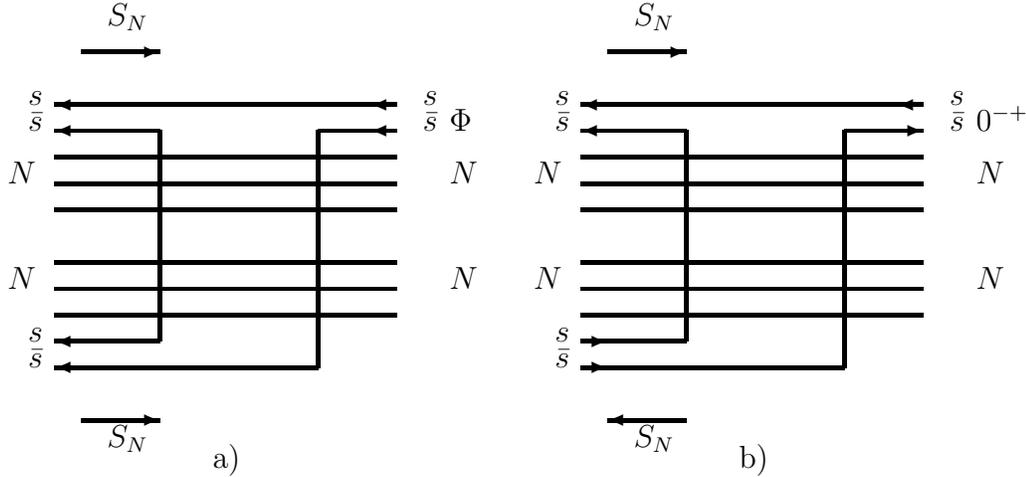
\begin{figure}[htb]
\setlength {\unitlength} {0.7mm} \thicklines
\linethickness{0.5mm}
\begin{picture}(180,70)(0,10)
\put(60,20){\vector(-1,0){50}}

\put(30,25){\vector(-1,0){20}}

\put(10,30){\line(1,0){65}}

\put(10,35){\line(1,0){65}}

\put(10,40){\line(1,0){65}}


\put(70,70){\vector(-1,0){60}}

\put(30,65){\vector(-1,0){20}}

\put(10,60){\line(1,0){65}}

\put(10,55){\line(1,0){65}}

\put(10,50){\line(1,0){65}}


\put(30,65){\line(0,-1){40}}

\put(60,20){\line(0,1){45}}


\put(75,70){\vector(-1,0){5}}

\put(75,65){\vector(-1,0){5}}

\put(60,65){\line(1,0){10}}

\put(75,35){\line(-1,0){20}}

\put(75,30){\line(-1,0){20}}


\put(40,1){a)}

\put(5,20){$\bar{s}$}

\put(5,25){$s$}

\put(5,65){$\bar{s}$}

\put(5,70){$s$}

\put(80,65){$\bar{s}$}

\put(80,70){$s$}

\put(85,65){$\Phi$}

\put(1,35){$N$}

\put(1,55){$N$}

\put(85,35){$N$}

\put(85,55){$N$}

\put(20,85){$S_{N}$}

\put(20,5){$S_{N}$}


\linethickness{0.5mm}

\put(15,10){\vector(1,0){15}}

\put(15,80){\vector(1,0){15}}



\put(110,20){\vector(1,0){5}}

\put(160,20){\line(-1,0){45}}

\put(110,25){\vector(1,0){5}}

\put(130,25){\line(-1,0){15}}

\put(110,30){\line(1,0){65}}

\put(110,35){\line(1,0){65}}

\put(110,40){\line(1,0){65}}


\put(170,70){\vector(-1,0){60}}

\put(130,65){\vector(-1,0){20}}

\put(110,60){\line(1,0){65}}

\put(110,55){\line(1,0){65}}

\put(110,50){\line(1,0){65}}


\put(130,65){\line(0,-1){40}}

\put(160,20){\line(0,1){45}}


\put(175,70){\vector(-1,0){5}}

\put(170,65){\vector(1,0){5}}

\put(160,65){\line(1,0){10}}

\put(175,35){\line(-1,0){20}}

\put(175,30){\line(-1,0){20}}


\put(140,1){b)}

\put(105,20){$\bar{s}$}

\put(105,25){$s$}

\put(105,65){$\bar{s}$}

\put(105,70){$s$}

\put(180,65){$\bar{s}$}

\put(180,70){$s$}

\put(185,65){$0^{-+}$}

\put(101,35){$N$}

\put(101,55){$N$}
\put(185,35){$N$}
\put(185,55){$N$}
\put(120,85){$S_{N}$}
\put(120,5){$S_{N}$}
\linethickness{0.5mm}
\put(130,10){\vector(-1,0){15}}
\put(115,80){\vector(1,0){15}}
\end{picture}
\vspace*{1cm}
\caption{Production of the \s~ mesons in $NN$ interaction
from the spin-triplet (a) and spin-singlet (b) states.
The arrows
show the direction of spins of the nucleons and strange quarks.}
\label{fig:8}
\end{figure}

        Due to the interaction it is possible that these pairs could
be shaken-out from the nucleon or strange quarks from different
nucleons
could participate in some rearrangement process similar to one shown in
Fig. \ref{fig:8}. Let us assume further that the quantum numbers of the
\s~pair is $J^{PC}=0^{++}$ (later we will explain this choice).

        Then the shake-out of such pairs will not create \f~
or tensor \ten~meson, but
a scalar strangeonium. The \s~ systems with other quantum numbers
(like \f~ or \ten)  should produce due to the process where
strange quarks from {\it both} nucleons are participating.

 An example of
this rearrangement diagram is shown in Fig.~\ref{fig:8}.
       If the nucleon spins are parallel (Fig.~\ref{fig:8}a),
then the spins of
the $\bar{s}$ and $s$ quarks in both nucleons are also parallel.
If the polarization of the strange quarks does not change during the
interaction, then the $\bar{s}$ and $s$ quarks will have parallel
spins in the final state. The total spin of \s~ quarks will be $S=1$ and
if their relative orbital momentum is $L=0$, it means that the
strangeonium has the \f~ quantum numbers, if $L=1$, it will correspond to
the creation of tensor strangeonium, $f'_2(1525)$.

        If the initial $NN$ state is a spin-singlet, the spins of
strange quarks in different nucleons are antiparallel and
the rearrangement diagrams like that in Fig.\ref{fig:8}b may lead
to the preferential formation of the \s~ system with total spin $S=0$. It means
that for $L=0$ one should expect an additional production of strangeonia with the pseudoscalar quantum numbers
$0^{-+}$.

        There are two important aspects of this scheme: the choice
of the quantum number of the \s~ pair in the nucleon and notion
that it is just rearrangement rather than shake-out processes are
responsible for the \s~ mesons production.

   In principle there are different possibilities for the quantum
numbers of the \s~ component in the nucleon wave function.
It may have, for instance,
pseudoscalar quantum numbers $J^{PC} = 0^{-+}$ or vector $J^{PC} = 1^{--}$ ones.
Then the relative angular momentum $j$ between
the \s~ and $uud$ cluster with $J^P = 1/2^+$ should be $j=1$. However, it
is also possible that
the \s~pair has quantum numbers of the vacuum $J^{PC} = 0^{++}$,
then $j=0$ to provide quantum numbers of proton. It is up to the experiment
to determine which of these possibilities are realized in nature.

If the nucleon \s~ pair has
quantum numbers of \f-meson it will lead to serious problems.
 In this case one might expect some additional
\f~ production due to the strangeness, stored in the nucleon.
This
quasi-\f~ pair could be easily shaken-out from the nucleon.
Then it is not clear how to explain the strong dependence of the
\f~ yield on quantum numbers of {\it both} nucleons, observed
at LEAR experiments.

Moreover, the shake-out of the \f~'s stored in the nucleon
should lead to an apparent violation  of the
OZI rule
in {\it all}
reactions of the \f~ production.

        Similar arguments were provided in \cite{Dov.90}, where
it was demonstrated that the experimental
data on the production of $\eta$ and $\eta'$ mesons exclude the $0^{-+}$
quantum numbers for the \s~ admixture in the nucleon wave function.

        In \cite{Alb.95} it was argued that the strange nucleon
sea may be negatively polarized due to the interaction of the light
valence quarks with the QCD vacuum. Due to the chiral dynamics
the interaction between quarks and antiquarks is most strong in
the pseudoscalar $J^{PC}=0^{-+}$ sector. This strong attraction
in the spin--singlet pseudoscalar channel between light valence
quark from the proton wave function and a strange antiquark from
the QCD vacuum will result in the spin of the strange antiquark
which will be aligned opposite to the spin of the light quark (and, finally,
opposite to the proton spin). As strange antiquark comes from the
vacuum, the corresponding strange quark to preserve the vacuum
quantum numbers $J^{PC}=0^{++}$ should also be aligned opposite to
the nucleon spin.

From the QCD sum rules analysis \cite{Iof.81},\cite{Rei.84}
it is known that the condensate of the strange
quarks in the vacuum is not small and is comparable with the condensate
of the light quarks:

\begin{equation}
< 0\left|\bar{s}s \right|0> = (0.8\pm0.1)< 0\left|\bar{q}q \right|0>,~~
q=(u,d)
\end{equation}

Thus, the density of $\bar{s}s$ pairs in the QCD vacuum is quite high and
one may expect that the effects of the polarized strange quarks in
the nucleon will
be also non-negligible.

        Therefore, we arrive to the picture of the negatively polarized
\s~ pair with the vacuum quantum numbers $^3P_0$.
These strange
quarks should not be considered like constituent quarks formed
some five quarks configuration of the nucleon. Rather they are
included in the components of a constituent quark.
It is important to stress that the \s~ pair with the $^3P_0$
quantum numbers  itself is not polarized  being a scalar.
That is a chiral non--perturbative interaction which selects only
one projection of the  total spin of the \s~ pair on the direction of
the nucleon spin.

        Since the quantum numbers of the \s~pair in nucleon
are fixed to be $^3P_0$, it is clear that the shake-out of this
state is not relevant for the \f~meson production. The rearrangement diagrams like those in
Fig.~\ref{fig:8} are responsible for an additional source of
\f~production. So on top of the \f~ production via standard reaction of  mixing
with light quarks there is an additional source of \s~ meson
production due to intrinsic nucleon strangeness. The rearrangement
nature of this mechanism implies that two nucleon should
participate in it. This means a dependence on quantum numbers of both
nucleons as well as appearance of some minimal momentum transfer from
which this additional mechanism becomes important.

It explains the $Q^2$ dependence of  ratio $R=\phi X/\omega X$ for different reactions of $\bar pp \to \phi (\omega) X$
annihilation at rest
shown in Fig. \ref{ozi}. Indeed, the largest OZI-violation has
been observed for the reactions with the largest momentum transfer to
\f. That is Pontecorvo reaction $\bar{p}d\to \phi n$ and $\bar pp \to \phi\gamma, \phi \pi$
processes. The kinematics of \ap~\an~ at rest restricts the
variation of momentum transfer. It is important to study the
dependence of
the violation of OZI rule on momentum transfer directly for \an~ in
flight.

The rearrangement nature of additional \f~ production allows to
make some interesting prediction concerning \an ~ into the $\phi \eta$ final state.
This channel was measured
by the OBELIX collaboration \cite{Nom.98} for
the $\bar pp$ annihilation at rest
in
liquid hydrogen, gas at NTP and at low pressure  of $5$ mbar.
The $\phi\eta$ final state has the same $J^{PC}$ as the
$\phi\pi^0$ final state. So, one may expect to see the same
selection rule as eqs.(\ref{rfi1})-(\ref{rfi2}) and suppose
that the \f~production in the low pressure sample will be
suppressed.
However, absolutely unexpectedly, the reverse trend is seen:
the yield of the $\bar{p}p \to \phi \eta$ channel grows
with decreasing of the target density.
The branching ratio for \an ~ from the $^1P_1$ state turns out to be
by
10 times higher than that of the $^3S_1$ state:
\begin{eqnarray}
B(\bar{p}p \to\phi\eta, ^3S_1) & = &(0.76\pm 0.31)\cdot10^{-4} \label{r1} \\
B(\bar{p}p\to\phi\eta, ^1P_1) & = &(7.72\pm 1.65)\cdot10^{-4}  \label{r2}
\end{eqnarray}

        Moreover, the Crystal Barrel measurements of annihilation
  in liquid give
the ratio of the phase space corrected branching ratios
\cite{Ams.98}:

\begin{equation}
R_\eta={B(\bar p p \rightarrow \phi \eta )
\over B( \bar p p \rightarrow \omega \eta)}= (4.6\pm 1.3)\cdot 10^{-3}
\label{Reta}
\end{equation}
in a perfect agreement with the OZI--rule prediction for the
 vector mesons (\ref{rfi}).

Polarized strangeness model treats these facts as demonstration
of the momentum transfer dependence: the momentum transfer in the
$\phi \eta$ reaction is too small for rearrangement diagrams start
working. So no OZI rule violation should be neither for \an~ from the S-wave
nor from the P-wave. The ratio $Y(\phi \eta)/Y(\omega \eta)$
should remain small in the P-wave. Therefore ten times increasing of the
$\omega \eta$ yield for \an~ from the P-wave is predicted.

       The  rearrangement nature of the additional strangeness
production implies strong consequences for the possible amount
of strange quarks in the nucleon. Indeed, let us estimate how many
additional \f~ are creating due to intrinsic nucleon strangeness.
Assuming the validity of the OZI rule and starting from branching
ratio $B.R. (\bar{p}p\to\omega \pi^0) =(63\pm4)\cdot 10^{-4}$ \cite{Nom.00} for the
\an ~ from the $^3S_1$ state, one may calculate the branching ratio of
$\bar{p} p \to \phi \pi^0$ reaction via normal OZI-allowed mixing.
It will give $B.R. (\phi \pi^0)_{OZI}  =  2.6\cdot 10^{-5}$, which
should be compared with the experimental value of
$B.R. (\phi \pi^0)_{exp} = (7.57\pm0.62)\cdot 10^{-4}$
\cite{Pra.98}.

Let us assume that all additional $\phi$'s are coming from the nucleon
intrinsic strangeness due to the two-nucleon rearrangement and try
to estimate (in a rather crude manner) the amount of strange
quarks admixture in the nucleon which is nedeed to create this
surplus of \f~.

The proton wave function is decomposing in two parts:
\begin{equation}
|p>  = a \sum^{\infty}_{X=0}|uud X>   + b \sum^{\infty}_{X=0}|uud \bar{s}s X>
\end{equation}
where $X$ stands for any number of gluons and light $\bar{q}q$
pairs. The normalization of the non-strange and strange parts of the wave function is $|a|^2 + |b|^2 = 1$
(neglecting the admixture of more than one \s~pair).
The amplitude of the \s~ meson production should be proportional
to $b^2$ due to the rearrangement nature of the process where strange
quarks from both nucleons should participate:

\begin{equation}
M(\bar{p}p \to \bar{s}s + X) \sim b^2 T(\bar{s}s)  \label{mss}
\end{equation}
where the factor $T(\bar{s}s) $ reflects the dependence on the initial
and final state interactions.

Let us assume that $T(\bar{s}s) \sim T(\bar{q}q)$. Then the ratio
R
of the {\it amplitudes} for the $\phi \pi$ and $ \omega \pi$ final
state is
\begin{equation}
R=\frac{M(\bar{p}p\to \phi \pi)}{M(\bar{p}p\to \omega \pi)}\sim
\frac{b^2}{1-b^2} \label{rm}
\end{equation}
Using experimental data for the additional strangeness production, one could
obtain that $R= 0.24\pm0.02$, it means that $b^2 = 0.20$.

        Of course, one should not consider this estimation
literally, as demonstration that the LEAR experiments found the 20\%
strangeness component in the nucleon. The approximation
$T(\bar{s}s) \sim T(\bar{q}q)$ is too crude and the reality may
differ on some factor. But what is clear: the rearrangement diagrams
mean the $b^2$ dependence of the \f~ production amplitude and degree
of the observed OZI violation (\ref{Rgamma})-(\ref{Rpitwo}) is so
high that indeed implies the contribution of the strange quarks
in the nucleon at the level of 10-20\%.

\section{Polarized strangeness model - experimental verification}

        The predictions of the polarized strangeness model have
been tested in different reactions. The detailed review of the present
status of the model is given in  \cite{Ell.99}.
Here I would like to comment only few recent experimental results:

$\bullet$ The Pontecorvo reactions

We already discussed that the ratio $R= Y(\bar{p} d\to \phi n)/Y(\bar{p} d\to\omega
n)$ shown in Fig. \ref{ozi} turns out to be unusually high. It is even greater than the corresponding ratio
for the annihilation on a free nucleon
$\bar{p} p  \to  \phi \pi^0 $ in liquid hydrogen and
twice as large as
the ratio  measured in hydrogen gas at NTP.
Usually, the Pontecorvo reactions are considered as
two-step processes \cite{Kon.89,Kon.98}. First,
two mesons are created in the
$\bar p$ annihilation on a single nucleon of the deuteron and then
one of them is absorbed by the
spectator nucleon. In this approach, the OZI violation in
the Pontecorvo reaction $\bar{p} d\to \phi n $ is simply a reflection of its
violation in the elementary act  $\bar{p} p  \to  \phi \pi^0 $.

        However recently the two-step model explanations are in
serious doubts after measurements by the Crystal Barrel collaboration
of the Pontecorvo reactions with open strangeness \cite{Abe.99}:
\begin{eqnarray}
  \bar{p} + d \to & \Lambda + K^0 \label{lk}\\
\bar{p} + d \to & \Sigma^0 + K^0 \label{sk}
\end{eqnarray}
It was found \cite{Abe.99}
that the yields of these reactions are practically
equal,\\
$R_{\Sigma,\Lambda}= Y(\Sigma K)/Y(\Lambda K) = 0.92\pm0.15$,
 in a sheer discrepancy with two-step models prediction \cite{Kon.89}
that the $\Sigma$ production (\ref{sk}) should be about 100 times less than
\la~
production. It was predicted \cite{Kon.98} that $R_{\Sigma,\Lambda}=0.012$.
This hierarchy appears naturally in the
two-step models due to the fact that the $\bar{K}N \to \Lambda X$ cross
section is larger than
the $\bar{K} N \to \Sigma X$ one.

The measured yields of  the reactions (\ref{lk})-(\ref{sk}) are
also at least by a factor 10 over the two-step
model prediction \cite{Kon.89}.

        Therefore, experiments on Pontecorvo reactions clearly indicate
opulent production of additional strangeness either in form of
\f~mesons or of $\Lambda K$ and $\Sigma K$ pairs.

$\bullet$ Longitudinal polarization of \la~ in DIS

It has been pointed out in \cite{Ek.95}
that the negative polarization of the strange sea should
lead to the negative longitudinal polarization of the \la~hyperons
formed
in the target fragmentation region in the lepton deep-inelastic scattering (DIS).
The idea is that
the polarization of the $s$ quark transfers to the final state polarization of the
\la~hyperon after an intermediate boson hits a valence quark of the
nucleon. Recent NOMAD data on $\nu N$ DIS \cite{NOM.00} have demonstrated that
the \la~ longitudinal polarization in the target fragmentation
region is indeed large and negative: $P(\Lambda)=-0.21\pm0.04\pm0.02$.
Remarkably, in the current fragmentation region this polarization
is significantly less: $P(\Lambda)=-0.09\pm0.06\pm0.03$. The measurements
of \la~ polarization with significantly large statistics is
planned at COMPASS experiment on $\mu N$ DIS.

$\bullet$  Polarization transfer to \la~ in $pp$ and \p

        The polarized strangeness model assumes an anticorrelation
  between spins of the proton and s-quarks. Then it is natural to predict
  a negative value for depolarization $D_{nn}$ measured in the
  \la~ production in
  polarized proton interactions $\vec{p}p \to \Lambda K^+ p$.
  Recent measurements  of DISTO collaboration \cite{Bal.98} indeed
 have confirmed this prediction. The same effect is expected for the
  depolarization of \la~ produced in \ap~ interactions with
  polarized protons $\bar{p} + \vec{p} \to \Lambda +
  \bar{\Lambda}$. However preliminary results of the PS 185
  experiment presented at this conference \cite{Pas.00} show that $D_{nn}$
  is quite small but the spin transfer to \al~$K_{nn}$ is unusually high
  and positive.

        There are versions of the polarized strangeness model
  where the spin of proton is indeed mainly transferred to
  \al~ rather than to \la. But in these modifications the
  $K_{nn}$ should be still negative. So, it is for the future to
  resolve this paradox.

        {\bf Acknowledgements:} I am extremely grateful to M.Alberg, J.Ellis, M.Karliner,
D.Kharzeev, A.Kotzinian, D.Naumov, K.Paschke and B.Popov for the
        numerous discussions.

\end{document}